Large magnetoresistance and magnetocaloric effect above 70 K in $Gd_2Co_2Al$, $Gd_2Co_2Ga$ and $Gd_7Rh_3$

Kausik Sengupta, Kartik K. Iyer and E.V. Sampathkumaran[*]
Tata Institute of Fundamental Research, Homi Bhabha Road, Colaba, Mumbai – 400005, India.

The electrical resistivity, magnetization and heat-capacity behavior of the Gd-based compounds, $Gd_2Co_2Al$, $Gd_2Co_2Ga$ and $Gd_7Rh_3$, ordering magnetically at $T_C$= 78 K, $T_C$= 76 K and $T_N$= 140 K have been investigated as a function of temperature and magnetic field. All these compounds are found to show large magnetoresistance (with a negative sign) in the paramagnetic state at rather high temperatures with the magnitude peaking at respective magnetic ordering temperatures. There is a corresponding behavior in the magnetocaloric effect as inferred from the entropy derived from these data.

PACS numbers: 73.43.Qt; 75.30.Sg; 75.20.En

In recent years, there is a lot of enthusiasm in studying Gd containing compounds, following the discovery of large magnetocaloric effect (MCE) in $Gd_5Si_2Ge_2$ by Pecharsky and Gschneidner [1] as well as large magnetoresistance (MR) and anomalous electron scattering effects [2-5] in the paramagnetic state in some of the Gd compounds. Thus, from the fundamental as well as applied science point of view, Gd compounds now present challenging situations. In this context, we report MR and MCE behavior for three Gd compounds, viz., $Gd_2Co_2Al$, $Gd_2Co_2Ga$ and $Gd_7Rh_3$. While $Gd_2Co_2Al$ and $Gd_2Co_2Ga$ form in the $Pr_2Ni_2Al$-type orthorhombic structure ordering ferromagnetically around 77 K [6], $Gd_7Rh_3$ crystallizes in $Th_7Fe_3$-type hexagonal structure ordering antiferromagnetically at 140 K [7]. We find that these materials exhibit large MCE in the vicinity of respective magnetic ordering temperatures ($T_o$). This gains importance considering a major emphasis in the current literature to discover materials with a large MCE above 20 K. In addition, the magnitude of MR is large (the sign being negative) peaking near $T_o$ qualitatively resembling the temperature dependence of MCE as though there is a relationship between these two properties.

The polycrystalline samples were prepared by arc melting stoichiometric amounts of the high purity (>99.9%) constituent elements together in an arc furnace in an atmosphere of argon. The ingots of $Gd_2Co_2Al$, $Gd_2Co_2Ga$ and $Gd_7Rh_3$ were homogenized in an evacuated sealed quartz tube at 870 C, 770 C and 300 C for 30, 30 and 2 days respectively. The samples were characterized to be single phase by x-ray diffraction. The electrical resistivity ($\rho$) and heat-capacity (C) measurements were performed in the temperature (T) interval 1.8 – 300 K in zero as well as in externally applied magnetic fields (H) up to 140 kOe with the help of a commercial (Quantum Design) Physical Property Measurements System. Magnetization (M) measurements were performed (4.2 – 300 K) as a function of temperature as well as H up to 120 kOe employing a vibrating sample magnetometer (Oxford instruments).

The results of $\rho$ measurements on $Gd_2Co_2Al$ and $Gd_2Co_2Ga$ as a function of temperature are shown in figures 1a and 1b. As expected, the temperature derivative of $\rho$ is positive in the paramagnetic state and there is a sudden drop in $\rho$ at the onset of magnetic ordering as reported previously [6]. As the magnetic field is applied, there is a negligible change in the $\rho$ values in the high temperature range (above 200 K and hence not shown in the form of a figure). However, as the temperature is lowered towards $T_C$, particularly below about $2T_C$, the values of $\rho$ are lower compared to those in zero field and the deviation peaks at $T_C$, as shown in figures 1c and 1d in the form of a plot of MR [$= ((\rho(H)-\rho(0))/ \rho(0)$]. We have also taken $\rho$ data as a function of H at fixed temperatures and it is distinctly clear from the plots in figures 1e and 1f that the trends are consistent with the observations made above. Thus, for instance, the curves in these figures for T near $T_C$ take highest values of MR. Thus, the magnitude of MR keeps increasing with decreasing temperature down to $T_C$ attaining rather large values even in the paramagnetic state. Thus, for instance, for H= 80 kOe, at 85 K (about 10 K higher than $T_C$), the value of MR falls in a range of 12 – 14% for both the compounds. As elaborated in the past [3,5], this finding, presumably arising from a novel spin scattering process, is interesting considering that the magnitude of MR is expected to be small in this temperature range in the paramagnetic state. Therefore, these compounds can be classified as metallic giant magnetoresistance (GMR) systems. A similar MR behavior has been reported for $Gd_7Rh_3$ as well [5] with the values being unusually large even at room temperature (in the paramagnetic state!) and since these MR results have been discussed at length in the earlier publication, we will not repeat presentation of these data in this paper for this compound.

Before we close the discussion on MR results, we would like to point out that there is a sign crossover of MR below 25 K with the magnitude of (positive) MR increasing with decreasing temperature, for $Gd_2Co_2Al$. Such an observation was made by us in the past in some ferromagnetic compounds, e.g., GdNi, possibly arising from interesting Fermi surface effects [9].

Considering current interests to discover materials for magnetic cooling, we have attempted to understand the MCE behavior of these compounds. The principle behind the application for this cooling process involves isothermal magnetization of a solid followed by adiabatic demagnetization. MCE results from the entropy change [defined here as $\Delta S = S(H)-S(0)$] in this process. One can obtain $\Delta S$ directly from the C data or from the magnetization data employing Maxwell's thermodynamic relation $(\partial S/\partial H)_T = (\partial M/\partial T)_H$ and it has been established that both the methods yield similar results [10]. We have therefore measured isothermal M up to 120 kOe at several temperatures at close intervals (about 5 K) in the vicinity of respective $T_o$. In addition, we have taken C data from 1.8 K to a desired temperature in zero as well as in the presence of one particular H to confirm the information obtained from M data as described above. It is also known [1] that the sign of $\Delta S$ carries information about the nature of the magnetic structure – that is, negative peak in $\Delta S$ is associated with ferromagnetism, whereas positive peak arises from antiferromagnetism. Corresponding peaks in $\Delta T$, the change in adiabatic temperature (defined as $T(H)-T(0)$), should be observed, however, with a sign opposite to that of $\Delta S$.

With this background, we now present the isothermal M behavior for all the three compounds in figure 2 for increasing H. Though we have taken the data at several temperatures, for the sake of clarity, we show the plots at selected temperatures only. In the case of $Gd_2Co_2Al$ and $Gd_2Co_2Ga$, as reported before, below $T_C$, there is a sudden rise of M below 10 kOe, beyond which there is a tendency for saturation and the saturation moment per Gd (extrapolated to zero H from high-field region) is about 1 to 1.5 $\mu_B$ less than that theoretically expected for Gd (7 $\mu_B$), thereby establishing that there is a moment induced on Co, which couples antiferromagnetically with the moments on Gd. Apparently this induced-moment is due to ferromagnetic ordering of Gd ions, as the effective moment obtained in the paramagnetic state (say, above 150 K) is typical of that expected for free Gd ions. On the other hand, in the case of $Gd_7Rh_3$, as reported earlier [5, 7], there is a metamagnetic transition around 60 kOe, and the saturation moment is higher (by about 1$\mu_B$) as though there is an induced-moment on Rh coupling ferromagnetically with the moment on Gd. The qualitative differences in M-H curves are actually reflected in MCE behavior as described now. The values of $\Delta S$ derived from these data are plotted in figure 3 for various values of H to enable a comparison of the MCE behavior. For the ternary compounds, the values of $\Delta S$ at their respective $T_o$ is noticeably large (about 5 J/mol K) for H= 50 kOe, whereas for $Gd_7Rh_3$, the corresponding values are negligible. However for higher fields, say beyond H= 70 kOe, $\Delta S$ becomes comparable (close to 8 J/mol K), thereby emphasizing the role played by metamagnetic transition in $Gd_7Rh_3$ to decide MCE. A peak in $\Delta S(T)$ at the respective $T_o$ is observed and the sign is consistent with ferromagnetic ordering (for $Gd_7Rh_3$ at very high fields). The finding of central importance is that the MCE as inferred from $\Delta S$ is moderately large at high fields in all these cases over 2 decades of temperature in the vicinity of $T_o$.

In order to support the conclusion arrived above, we obtained C in zero and in the presence of a magnetic field. To our knowledge, this is the first report on the C behavior of all these compounds and therefore it is worth making some comments on C(T) (see figure 4). There is a sharp peak at the onset of magnetic ordering in all these cases. Zero-field C of $Gd_7Rh_3$ is found to obey $T^3$-dependence at low temperatures (below 14 K) expected for antiferromagnets. However, in the case of the binary Gd compounds under discussion, the $T^{3/2}$ expected for simple ferromagnets is not observed and we believe that the functional form could be more complex due to the antiferromagnetically-coupled induced-Co-moment. However, the peak gets broadened in the presence of a relatively low field of 30 kOe for the ternary compounds with a reduction in peak-height, whereas for $Gd_7Rh_3$ one has to apply a higher field (which can induce the metamagnetic transition, say, 80 kOe) to see this broadening. In the case of the ternary compounds, the peak actually shifts to a marginally higher temperature typical of ferromagnets, say for H= 30 kOe. In the case of $Gd_7Rh_3$, there is actually a downward shift (say, by about 7 K, see Fig. 4) of $T_0$ even for initial applications of H= 50 kOe, as expected for antiferromagnets; however, the peak position interestingly shifts marginally upwards for a field as high as 80 kOe. This means that, for $Gd_7Rh_3$, the exchange interaction strength following metamagnetic transition is not depressed though the sign changes. We have derived $\Delta S(T)$ on the basis of the knowledge of C in zero field and in the presence of H and plotted in figure 3 along with the curves determined from M data. One can see that there is a fairly good agreement between the results obtained from these two experimental methods within the limits of experimental error, thereby providing confidence in our analysis. $\Delta T(T)$ behavior as determined from the C data are also plotted as insets in figure 4 to get an idea of the change in temperature following the adiabatic process.

To summarise, we have reported large magnetoresistance over a wide temperature range in the paramagnetic state of three Gd-based compounds in a convenient temperature range, that is, easily accessible with liquid nitrogen. In addition, the MCE as measured by $\Delta S$ and $\Delta T$ also exhibit a behavior similar to that of magnetoresistance, thereby bringing out that a common mechanism involving spin controls these two different phenomena.

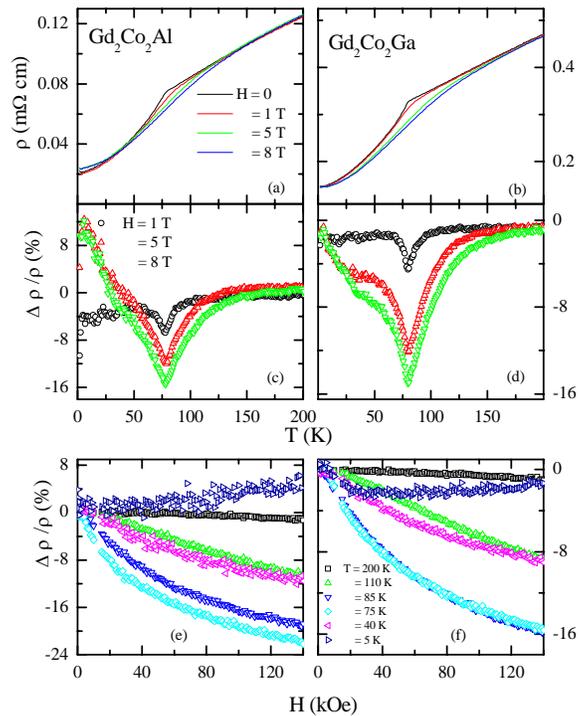

Figure 1:

Electrical resistivity (ρ) as a function of temperature (below 200 K) in the absence and in the presence of externally applied magnetic fields (H) for (a) $Gd_2Co_2Al$ and (b) $Gd_2Co_2Ga$. The magnetoresistance, MR, defined as $(\rho(H)-\rho(0))/\rho(0)$, derived from these data are plotted in (c) and (d) respectively. MR obtained as a function of H at fixed temperatures are shown in (e) and (f) respectively.

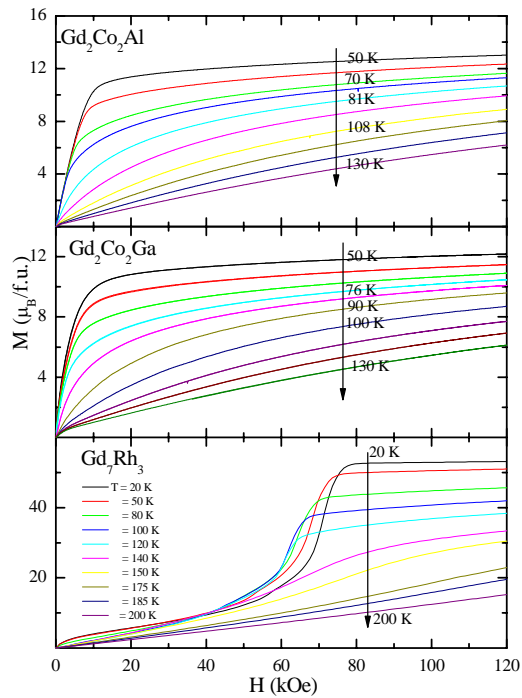

Figure 2:

Isothermal magnetization (M) as a function of increasing externally applied magnetic fields for $Gd_2Co_2Al$, $Gd_2Co_2Ga$ and $Gd_7Rh_3$ at many temperatures. For the sake of clarity, only some temperatures are marked for the ternary compounds.

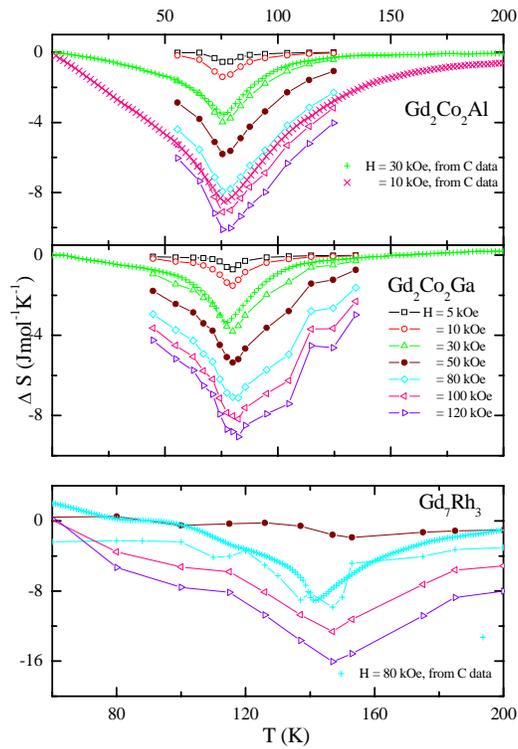

Figure 3:

The entropy change, ΔS = S(H)-S(0), as a function of temperature, obtained from isothermal magnetization data for different values of H for $Gd_2Co_2Al$, $Gd_2Co_2Ga$ and $Gd_7Rh_3$. The curves obtained from the heat-capacity data (shown in figure 4) for specified fields are also shown for comparison.

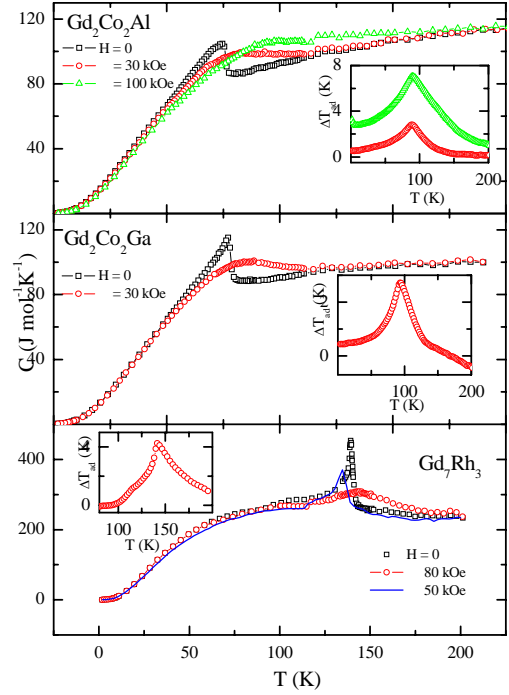

Figure 4:

Heat-capacity (C) as a function of temperature in zero and in the presence of external fields for $Gd_2Co_2Al$, $Gd_2Co_2Ga$ and $Gd_7Rh_3$. Adiabatic change in temperature, $\Delta T = T(H)-T(0)$, derived from the C data, are shown in the insets.